\documentstyle[12pt]{article}

\catcode`\@=11

\def\@fnsymbol#1{\ifcase#1\or * \or \dagger\or \ddagger\or 
\mathchar "278\or \mathchar "27B\or \|\or **\or \dagger\dagger
\or \ddagger\ddagger \else\@ctrerr\fi\relax}%

\def\title{%
\vspace{0.5cm}\vspace{4ex}
\bgroup
\obeylines
\large\boldmath \bf\begin{center}
}
\def\endtitle{\end{center}\vskip1sp\egroup}
\def\author#1{\begingroup\center #1 \endcenter\endgroup}

\def\keywords#1{\par\noindent{\bf KEYWORDS}: #1}
\def\addcontentsline#1#2#3{\relax}
\def\fnum@figure{Figure \thefigure}
\def\fnum@table{Table \thetable}
\newcounter{figcaption}
\def\thefigcaption{\arabic{figcaption}}
\def\fnum@figcaption{{\bf Fig. \thefigcaption :}}
\def\figcaption{%
{\parindent 0pt \bf Figure Captions} \par \vskip 10pt
\list{\fnum@figcaption}
{\leftmargin 5em \labelwidth\leftmargin\advance\labelwidth-\labelsep
\def\makelabel##1{##1\hfil} \usecounter{figcaption}}%
}

\def\cite{\@ifnextchar [{\@tempswatrue\@citex}{\@tempswafalse\@citex[]}}
\def\@citex[#1]#2{\if@filesw\immediate\write\@auxout{\string\citation{#2}}\fi
\def\@citea{}\@cite{\@for\@citeb:=#2\do
{\if-\@citeb \mbox{-}\def\@citea{}
\else
\@citea\def\@citea{,\penalty\@m}\@ifundefined
{b@\@citeb}{{\bf ?}\@warning
{Citation `\@citeb' on page \thepage \space undefined}}%
\hbox{\csname b@\@citeb\endcsname}
\fi}}{#1}}
\newfont{\scrptrm}{cmr8}
\def\@cite#1#2{${}^{\scrptrm {#1\if@tempswa , #2\fi})}$}

\def\rcite{\@ifnextchar [{\@tempswatrue\@rcitex}{\@tempswafalse\@rcitex[]}}
\def\@rcitex[#1]#2{\if@filesw\immediate\write\@auxout{\string\citation{#2}}\fi
\def\@rcitea{}\@rcite{\@for\@rciteb:=#2\do
{\if-\@rciteb \mbox{-}\def\@rcitea{}%
\else
\@rcitea\def\@rcitea{,\penalty\@m}\@ifundefined
{b@\@rciteb}{{\bf ?}\@warning
{Citation `\@rciteb' on page \thepage \space undefined}}%
\hbox{\csname b@\@rciteb\endcsname}%
\fi}}{#1}}
\def\@rcite#1#2{{#1\if@tempswa , #2\fi}}
\def\refcite{\@ifnextchar [{\@tempswatrue\@refcitex}
{\@tempswafalse\@refcitex[]}}
\def\@refcitex[#1]#2{
\if@filesw\immediate\write\@auxout{\string\citation{#2}}\fi
\def\@citea{}\@refcite{\@for\@citeb:=#2\do
{\if-\@citeb -\def\@citea{}
\else
\@citea\def\@citea{,\penalty\@m}\@ifundefined
{b@\@citeb}{{\bf ?}\@warning
{Citation `\@citeb' on page \thepage \space undefined}}%
\hbox{\csname b@\@citeb\endcsname}
\fi}}{#1}}
\def\@refcite#1#2{[{#1\if@tempswa , #2\fi}]}

\newcommand{\rmc}{{\rm c}}
\newcommand{\rmd}{{\rm d}}
\newcommand{\rme}{{\rm e}}
\newcommand{\rmi}{{\rm i}}

\newcommand{\Ham}{{\cal H}}

\newcommand{\dags}{^\dagger}

\def\braket#1{\left\langle#1\right\rangle}

\def\brakets#1{\langle#1\rangle}

\def\simle{\mathrel{\mathpalette\@versim<}}   
\def\simge{\mathrel{\mathpalette\@versim>}}   
\def\@versim#1#2{\lower2.5pt\vbox{\baselineskip0pt \lineskip-.5pt
   \ialign{$\m@th#1\hfil##\hfil$\crcr#2\crcr\sim\crcr}}}

\newcommand{\bequ}{ \begin{equation} }
\newcommand{\eequ}{ \end{equation} }
\newcommand{\barr}{ \begin{array} }
\newcommand{\earr}{ \end{array} }
\newcommand{\beqarr}{ \begin{eqnarray} }
\newcommand{\eeqarr}{ \end{eqnarray} }

\newcommand{\baralpha}{ \begin{eqnal} \beqarr}
\newcommand{\earalpha}{ \eeqarr \end{eqnal}}

\catcode`\@=12

\def\MFreq{\rmi \omega_n}

\def\vecn{\vec m}

\def\LSMO{La$_{1-x}$Sr$_x$MnO$_3$}

\def\Neel{N\`eel}

\begin{document}

\begin{title}
Magnetic  Transition Temperature
of (La,Sr)MnO$_3$
\end{title} 

\author{Nobuo {\sc Furukawa}}

\begin{instit}
  Institute for Solid State Physics,\\
  University of Tokyo, Roppongi 7-22-1,\\
  Minato-ku, Tokyo 106
\end{instit}

\begin{abstract}
Using the Kondo lattice model with classical spins 
in infinite dimension,
 magnetic phase transition in  
the perovskite-type $3d$ transition-metal oxide {\LSMO}
is theoretically studied.
On the Bethe lattice, the self-consistency equations
are solved exactly.
Curie temperatures at the region of double-exchange
ferromagnetism $0.1 \simle x \simle 0.25$ as well as
the {\Neel} temperature at $x=0$ are well reproduced quantitatively.
Pressure effect on the Curie temperature is also discussed.
\end{abstract}  

\vfil

\keywords{Transition-metal oxide, manganese oxide, 
double-exchange ferromagnetism,
Kondo lattice model, infinite dimensions}

\pagebreak

After the discovery of high $T_\rmc$ oxides, 
the physics of strongly correlated 
$3d$ transition-metal oxide compounds have been revisited
extensively. 
One of such materials is the 
perovskite-type manganese oxides {\LSMO}.
The most prominent feature of the system
is  the giant magnetoresistance (GMR) 
with negative sign.
The system is a double-exchange ferromagnet\cite{Zener51,deGennes60}
 at $ x \simge 0.1$, while
the antiferromagnetic phase with spin canting 
is observed at $x \sim  0$.
For the detail of the
 phase diagram and the GMR in {\LSMO}
and related materials,
the readers are referred to {\em e.g.}\ 
refs.~\rcite{Tokura94} and \rcite{Urushibara9x}, and those cited therein.

In {\LSMO}, $3d$ electrons are considered to form
both localized spins with $S=3/2$ in the $t_{2\rm g}$ orbitals
 and itinerant electrons in the $e_{\rm g}$ orbitals  which are
coupled  each other by Hund's ferromagnetic interaction.
The bandwidth of the itinerant electron is estimated to be 
$W \sim 1{\rm eV}$ from the recent band calculation,\cite{Hamada94x}\ 
 while the Hund's coupling is considered to be
larger than the bandwidth.
As a model Hamiltonian of this system, the Kondo lattice model (KLM)
with  $S=3/2$ and  the ferromagnetic exchange
 coupling 
 has been proposed.\cite{Kubo72}\ \ 

{}From the strong coupling limit of the above model,
Curie temperature $T_\rmc$ has been
studied.\cite{deGennes60,Kubo72}\ \ 
Using a mean-field type  treatment, Curie temperature is estimated to be
$T_\rmc \sim W$, where $W$ is the bandwidth of the conduction electron.
This result is roughly understood as follows; the transition temperature
is determined from the competition between the gain of kinetic energy
in the ferromagnetic state and the gain of entropy in
the paramagnetic state. 

Quantitatively, however, the mean-field approach  fails to explain
the experimental value of $T_\rmc$,
{\em e.g.} $T_\rmc \sim 300 {\rm K}$ at $x \sim 0.2$
which is  a few orders of magnitude lower than the theoretical estimate.
The discrepancy seems to be due to the fact that the mean-field treatment
poorly describes the electronic state in the paramagnetic phase
so that kinetic energy is not
estimated properly.
In the paramagnetic phase,  thermal spin fluctuations 
 have to be taken into account appropriately
 because the system is in a strongly correlated regime.
Recently, Millis {\em et.al.} has made 
an analysis on $T_\rmc$ 
based on the calculation of
the spin stiffness at the ground state.\cite{Millis9x}\ \ 
Their result also overestimates $T_\rmc$.
Since the changes in the electronic state from the
ferromagnetic ground state to the paramagnetic state
should be nontrivial and  drastic, 
 ground state properties may not reproduce the transition temperature.
Because the GMR is most prominently observed in the vicinity of
the Curie temperature, a theoretical approach which is able to
predict $T_\rmc $ with accuracy is demanded also from the standpoint of
a material designing and applications.

The author has shown in his previous 
work\cite{Tokura94,Furukawa94,Furukawa95a} 
that 
the KLM in infinite dimensional limit $D=\infty$ and infinite high-spin
limit $S=\infty$ reproduces the transport properties of 
{\LSMO} quantitatively.
Within these limits,
Green's functions are obtained exactly even in the paramagnetic state.
Therefore, a proper treatment is possibly performed to obtain
magnetic instabilities in the paramagnetic phase.
In this paper, we study the magnetic phase transition of the
above model. One of the aims is to examine, within this model,
 whether it
is possible to explain the magnitude of the
Curie temperature as well as its doping dependence in a consistent way.

Thus we investigate the KLM at $D=\infty$ and $S=\infty$.
The Hamiltonian is described as
\bequ
  \Ham = 
  - t \sum_{<ij>,\sigma}
        \left(  c_{i\sigma}\dags c_{j\sigma} + h.c. \right)
    -J \sum_i \vec \sigma_i \cdot \vec m_i,
    \label{HamSinfty}
\eequ
where $ \vec m_i = (m_i{}^x, m_i{}^y, m_i{}^z)$ and $|\vec m|^2 = 1$.
We consider the Bethe lattice in the infinite coordination number
limit.
{}From the nature of the Bethe lattice that lattice points are
divided into two sublattices, it is possible to study  magnetic phases
with ferromagnetic and antiferromagnetic 
order parameters.
The density of states (DOS) is given in the semi-circular form 
$N_0(\varepsilon) = (2/\pi W) \sqrt{ 1 - (\varepsilon/W)^2}$,
where $W$ is the bandwidth.
Here we see another advantage of the Bethe lattice that the DOS
has similar properties to that of the $D=3$ systems, namely
the existence of a band edge and the shape of the DOS at the edge.
The above properties may be important 
in the case of making comparisons between
the theoretical results and the experimental data in the
 low hole concentration region.

The infinite-dimensional model is mapped to
that of the corresponding single-cell model.
The partition function is written in a form
\bequ
  Z = \int\!\!\int \rmd\Omega_{A}\rmd\Omega_{B} \ 
  Z_{\rm f}(\vecn_A,\vecn_B),
	\label{defPrtFun}
\eequ
where integration is performed with respect to the orientation
of local spins.
Using  the Weiss fields  $\tilde G_{0\alpha}(\MFreq) $
which describe the dynamical motions
of electrons 
 for $\alpha=A,B$ sublattice sites, 
the fermion trace is given by
\bequ
  Z_{\rm f}(\vecn_A,\vecn_B) = \!\!
 \prod_{\alpha=A,B} \!\!4 \exp \left( \sum_n \log \det \left[ 
       \frac{(\tilde G_{0\alpha}^{-1} + J\vecn_\alpha\vec\sigma)}
	    {\MFreq} 
    \right] \rme^{\MFreq 0_{+}} \right).
  \label{defZf}
\eequ
The Green's function is calculated exactly as
\bequ
  \tilde G_\alpha(\MFreq) = 
     \int\!\!\int \rmd\Omega_{A}\rmd\Omega_{B} \ 
           \frac{Z_{\rm f}(\vecn_A,\vecn_B)}{Z}
             \left(  \tilde G_{0\alpha}^{-1}(\MFreq) +
                    J \vecn_\alpha \vec \sigma \right)^{-1}.
\eequ
On the Bethe lattice with two-sublattice symmetry, the mapping
relation gives\cite{Rozenberg94}
\beqarr
   \tilde G_{0A}{}^{-1}(\MFreq) &=&
		 \MFreq + \mu - \tilde G_B ( \MFreq) /4 ,
		\nonumber\\
   \tilde G_{0B}{}^{-1}(\MFreq) &=&
		 \MFreq + \mu -\tilde  G_A ( \MFreq)/4 .
	\label{defSCCG}
\eeqarr
Self-consistency equations 
(\ref{defPrtFun})-(\ref{defSCCG})
are mainly solved in a numerical way.
At $J/W \gg 1$, 
Green's functions near the
magnetic transition temperature are also studied analytically.

Sublattice magnetization of the local spins 
$\brakets{\vec m_\alpha} $ is obtained from
\bequ
   \brakets{\vec m_\alpha} =
     \int\!\!\int \rmd\Omega_{A}\rmd\Omega_{B} \ 
          \vecn_\alpha  \frac{Z_{\rm f}(\vecn_A,\vecn_B)}{Z}.
 	\label{defMalpha}
\eequ
Then,
magnetic transition temperatures are 
determined as a function of Hund's coupling $J$ 
and chemical potential $\mu$ in the unit of $W$.
The carrier electron number for
{\LSMO} is nominally considered to be $n=1-x$. Hereafter we
use the hole picture so that the hole concentration is 
expressed by $\brakets{x} = 1 - \brakets{n}$.
In this paper, we restrict ourselves to 
$0 \le \brakets{x} \simle 0.5$, {\em i.e.} from half-filling to
quarter-filling.

In Fig.~\ref{FigTcJdep}, we show $T_\rmc/W$ as a function of
$J/W$ at various filling. Here, for simplicity of the
calculation,  chemical potentials are
systematically chosen to be  $\mu(J) = -J + \delta\mu$
with $\delta\mu/W=0$, $ 0.25$ and $ 0.33$
so that the carrier numbers at $T_\rmc$ become
$\braket{x} \simeq 0.5$, $0.3$ and $0.2$, respectively, with
errors $\simle \pm 0.03$ at $J/W \ge 4$. The curves in the figure
are the results of the  $(J/W)^{-1}$ expansion at $J/W \gg 1$ in the form
\bequ
	T_\rmc / W = \tilde T \left[ 1 - \tilde J (J / W)^{-1} \right].
    \label{defTcFitFun}
\eequ
We see $T_\rmc \propto W$ at $J/W \to\infty$.
Dimensionless constants
are given by $\tilde T \simeq 0.044$, $0.039$ and $0.035$
while $\tilde J \simeq 0.49$, $0.89$ and $1.12$
for $\braket{x} \sim 0.5$, $0.3$ and $0.2$, respectively.
The fitting curves seems to reproduce results 
also at $J/W\sim 1$ surprisingly well.
The quantity $\tilde T $ is the Curie temperature in the scale of $W$
at
$J/W\to\infty$, while $\tilde J$ may be interpreted as the
magnitude of the instability of the ferromagnetic order
 upon decreasement of $J/W$. 
{}From $\tilde T \sim 0.04$ and $W\sim 10^4{\rm K}$, we see that
the Curie temperature in {\LSMO} is roughly
explained within this model in the strong coupling limit.
The tendency that
$\tilde T$ ($\tilde J$) increases (decreases) as holes are doped
is easily understood 
 because the stability of the ferromagnetic order
should be  enhanced by hole doping.
 
Now, we precisely 
compare the doping dependence of  $T_\rmc$
with the experimental data. Calculations are performed 
at $J/W=4$ which is the value that explains the
magnetoresistance phenomena 
 successfully,\cite{Tokura94} as well as at $J/W=\infty$.
In Fig.~\ref{FigTcJ4}, we show  $T_\rmc$ as a function of 
 $x$ together with the
experimental data from ref.~\rcite{Urushibara9x}. 
Here, the bandwidth  is scaled 
so that $T_\rmc$ at $x=0.15$  reproduces the experimental result;
we then have $W=1.05{\rm eV}$ for $J/W=4$, which is 
a moderate value.
Thus we see from the above result that the experimental data are
quantitatively reproduced very well at $x \simle 0.25$.
Discrepancy on the doping dependence of $T_\rmc$ is
observed at $x \simge 0.3$, which may be interpreted as
a sign of a
 crossover of the system from the strong coupling regime
to a weak coupling limit.

Next, we calculate the pressure effect on $T_\rmc$.
In   {\LSMO} at
 the  ferromagnetic metal region  
$x \simge 0.15$,
increase of $T_\rmc$ is observed under pressure.\cite{Moritomo95x}\ \ 
The pressure coefficient $\rmd \log T_\rmc / \rmd P$ is positive,
 and it decreases systematically
 as the hole concentration is increased. 
For the theoretical treatment,
 we make following assumptions; the bandwidth 
increases as pressure is applied due to the increase of
 overlaps between neighboring orbitals,
while the intra-atomic Hund's coupling is not affected.
Then,  we have
\bequ
  \frac{\rmd \log T_\rmc}{\rmd P} = \frac{\rmd \log W}{\rmd P}
     \frac{\partial \log T_\rmc}{\partial \log W},
\eequ
where $ \rmd \log W / \rmd P > 0 $ from the assumption.
Using eq.~(\ref{defTcFitFun}) at $J \gg \tilde JW$, we have
\bequ
       \frac{\partial \log T_\rmc}{\partial \log W} \simeq
      1 - \tilde J (J/W)^{-1},
\eequ
so that a positive pressure coefficient is 
derived in the strong coupling region.
We have shown above that $\tilde J$ decreases as $x$ is increased.
Then, the pressure coefficient of $T_\rmc$ should increase by hole doping
if $J/W$ is fixed, which is contrary to the experimental result.
 In order to explain the experiment
 within the present model and the assumptions mentioned above,
 we must have the effective decrease in 
$J/W$ as the hole concentration is increased.
This is another implication that the system undergoes the crossover
to the weak coupling region by hole doping.

In  Fig.~\ref{FigPDx0}, the {\Neel} temperature $T_{\rm N}$
 at $x=0$ is shown.
At $J \gg W$, the expansion with respect to $(J/W)^{-1}$ gives
an analytical formula
\bequ
   T_{\rm N} / W  = (1/24)\cdot (J/W)^{-1}.
\eequ 
The {\Neel} temperature in LaMnO$_3$, 
$T_{\rm N} \simeq 140{\rm K}$,\cite{Urushibara9x}\ 
is also well explained, {\em e.g.}
if we set $J=4W$ and $W \simeq 1.2{\rm eV}$.
Upon hole doping, the model in the low temperature regime
shows a first order transition from the {\Neel} state directly to
the ferromagnetic state. Canted {\Neel} state which is
observed in {\LSMO} at $x \simle 0.1$ is not found
within this model at $J/W \simge 2$.
It must be noted that the present approach does not take into account the
magnetic phase with incommensurate wave numbers such as spiral states
or conical states.
In order to study the magnetic structure
 at $x \simle 0.1$,
 it seems to be necessary to treat the model in a realistic
lattice structure. Effects from the orbital degeneracy in $e_{\rm g}$ bands
and its Jahn-Teller splitting may also be essential to explain
the experimental results in that region. 

To summarize, we have calculated the KLM in
the limit $S=\infty$ and $D=\infty$ on the Bethe lattice. 
Magnetic transition temperature is obtained exactly
as a function of interaction strength  and hole concentration.
Comparison with the experimental data in {\LSMO} is made.
At $J/W\simeq 4$ and $W \simeq 1{\rm eV}$, the model consistently
explains the {\Neel} temperature at $x=0$ and Curie
temperatures at $0.1 \simle x \simle 0.25$ in a  quantitative way.

The author would like to thank Y. Tokura and T. Arima
for fruitful discussions and comments.
The numerical calculation is partially performed on the
 FACOM VPP500 at the Supercomputer Center, 
Inst.\ for Solid State Phys., Univ.\ of Tokyo.

\begin{figcaption}

\item Curie temperature $T_\rmc /W$ as a function of $J/W$.
\label{FigTcJdep}

\item Curie temperature $T_\rmc$ at $J/W=4$ 
and $J/W=\infty$  as a function of $x$.
Experimental data in {\LSMO} are from
ref.~\rcite{Urushibara9x}. Temperature is scaled in the unit of Kelvin.
\label{FigTcJ4}

\item {\Neel} temperature at $x=0$. The line in the figure is
the result from $(J/W)^{-1}$ expansion.
\label{FigPDx0}

\end{figcaption}
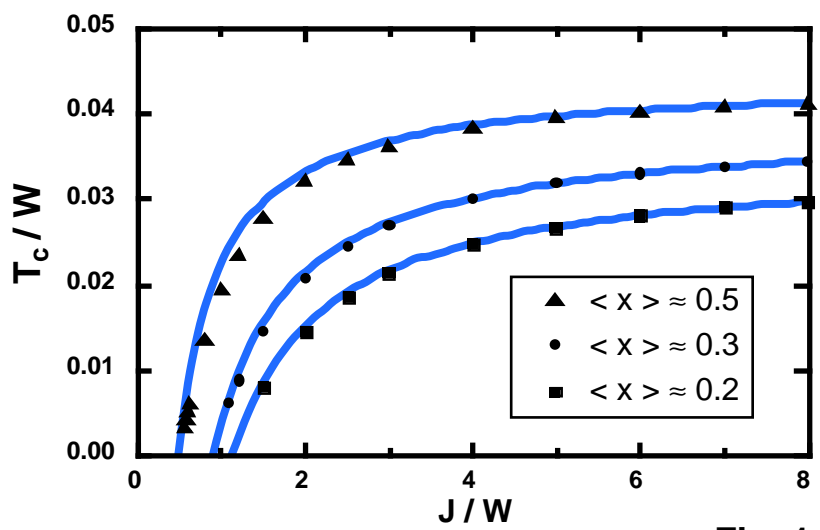
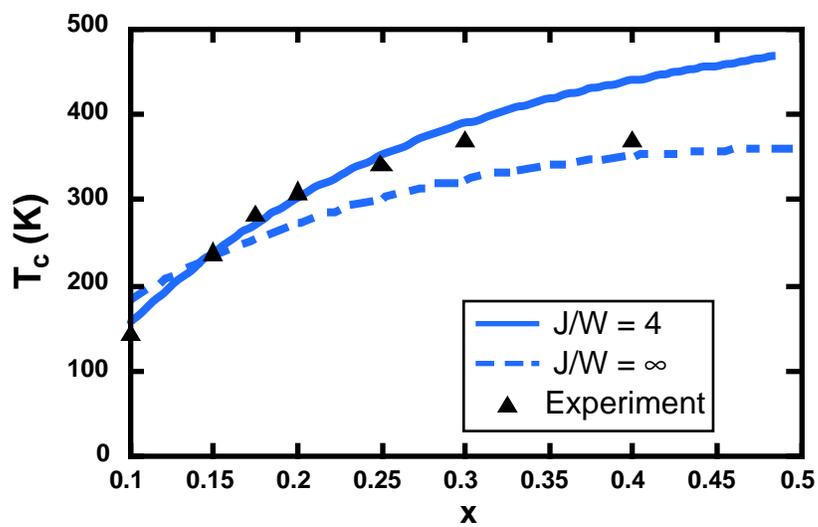
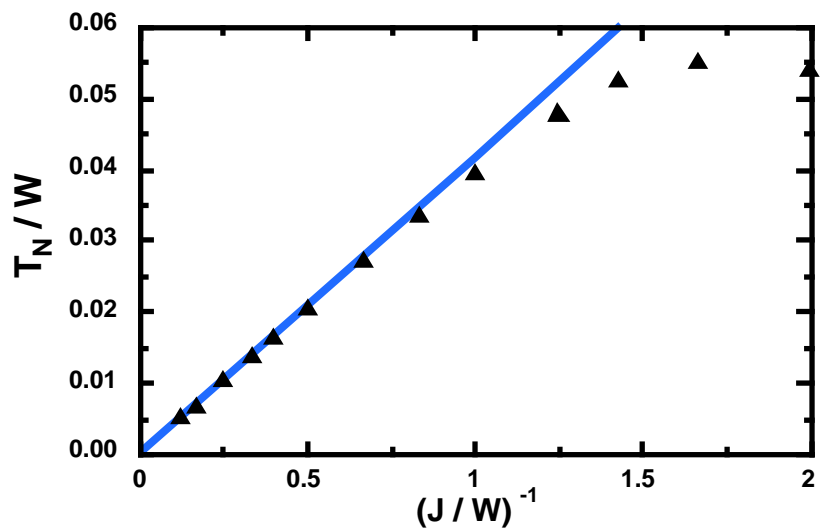
\end{document}